\documentclass[aps,prl,superscriptaddress,twocolumn]{revtex4-1}% Physical Review B

\usepackage{bm, amssymb}
\usepackage{dcolumn}
\usepackage{graphicx}
\usepackage[dvips]{epsfig}
\usepackage{xcolor}
\usepackage[T1]{fontenc}

\usepackage{bm, amssymb}
\usepackage{amsmath}
\usepackage{dcolumn}

\begin{document}
	
	\title{Pressure-induced collapse of ferromagnetism in Nickel}

	\author{A. Ahad}
	\thanks{aahad39@myamu.ac.in}
	\address{Department of Physics, Aligarh Muslim University, Aligarh 202002, India}
	\author{M. S. Bahramy}
	\thanks{m.saeed.bahramy@manchester.ac.uk}
	\address{Department of Physics and Astronomy,The University of Manchester, Manchester M13 9PL, United Kingdom}

\date{\today}

\begin{abstract}
Transition metals, Fe, Co and Ni, are the canonical systems for studying the effect of external perturbations on ferromagnetism. Among these, Ni stands out as it undergoes no structural phase transition under pressure. Here we have investigated the long-debated issue of pressure-induced magnetisation drop in Ni from first-principles. Our calculations confirm an abrupt quenching of magnetisation at high pressures, not associated with any structural phase transition. We find that the pressure substantially enhances the crystal field splitting of Ni-$3d$ orbitals, driving the system towards a new metallic phase violating the Stoner Criterion for ferromagnetic ordering.  Analysing the charge populations in each spin channel, we show that the next nearest neighbour interactions play a crucial role in quenching ferromagnetic ordering in Ni and materials alike.  
\end{abstract}

\maketitle

In transition metals such as Fe, Co and Ni, the strong exchange interaction between the valence shell 3$d$ electrons lead to itinerant ferromagnetic (FM) ordering with a sizable saturated magnetic moment $M_s$. External perturbations like pressure ($P$),  temperature and magnetic field enable manipulating this ordering and its underlying mechanisms. Among these stimuli, the application of $P$ is particularly of great utility, as it allows studying the interplay of spin, orbital and lattice degrees of freedom of electron under controlled conditions~\cite{Mao2018}. Accordingly, extensive experimental and theoretical studies have already been performed to understand the collective behaviour of $3d$ electrons in such cross-correlated systems. 

In general, $P$ tends to decrease the interatomic distances, thereby enhancing the overlap of the atomic wave functions~\cite{Mor1988} which itself results in band broadening and promotion of electrons to higher energy states, ultimately increasing the kinetic energy of electrons \textit{i.e.}, delocalization~\cite{Ser2013}. More specifically speaking, broadening reduces the density of states (DOS) at the Fermi level ($E_F$), leading to a violation of the Stoner criterion (SC) for spontaneous magnetism~\cite{Moh2006}, \textit{i.e.} $U \text{DOS}(E_F) > 1$, where $U$ is the Hubbard parameter. In the case of Fe and Co, such a collapse of FM ordering occurs at $\sim$14 and $\sim$100 GPa, respectively~\cite{Nic1963,Tor2011}. On the other hand, Ni?which has a close-packed face-centred cubic (FCC) structure with a relatively small $M_s=0.6 \mu_B$/Ni? can survive up to extremely high $P$s~\cite{Ser2013,Moh2010}. Previous \textit{ab-initio} calculations have suggested a monotonic decrease of $M_s$ in Ni for $P$s up to 100 GPa~\cite{Xie2000}. Under high $P$s, Ni tends to become non-magnetic but remains metallic. High-pressure X-ray magnetic circular dichroism (XMCD) experiments also have indicated~\cite{Iot2007} a weak decrease in K-edge signal in Ni, representing its magnetization, even when the volume ($V$) compresses by 80\% (\textit{i.e.} $V/V_0=0.8$, with $V_0$ being the volume at the ambient pressure). At higher compressions, the K-edge signal exhibit a more significant suppression, suggesting a phase transition to a non-magnetic state at $P=250$ GPa. This observation is, however, contradicted by a similar experiment~\cite{Tor2011b} showing the resistance of the FM phase for volume compressions up to $V/V_0=0.66$ ($P=200$ GPa). Using density functional theory (DFT) calculations, the authors of the latter work have claimed that the K-edge signal stems from the $4p$ orbital moment rather than the total spin moment and speculated that the non-magnetic state could only occur above 400 GPa. This claim is further supported by a Nuclear Forward Scattering experiment, demonstrating the survival of the FM phase up to 260 GPa~\cite{Ser2013}.  

\begin{figure}[t]
	\centering
	\includegraphics[width=0.45\textwidth]{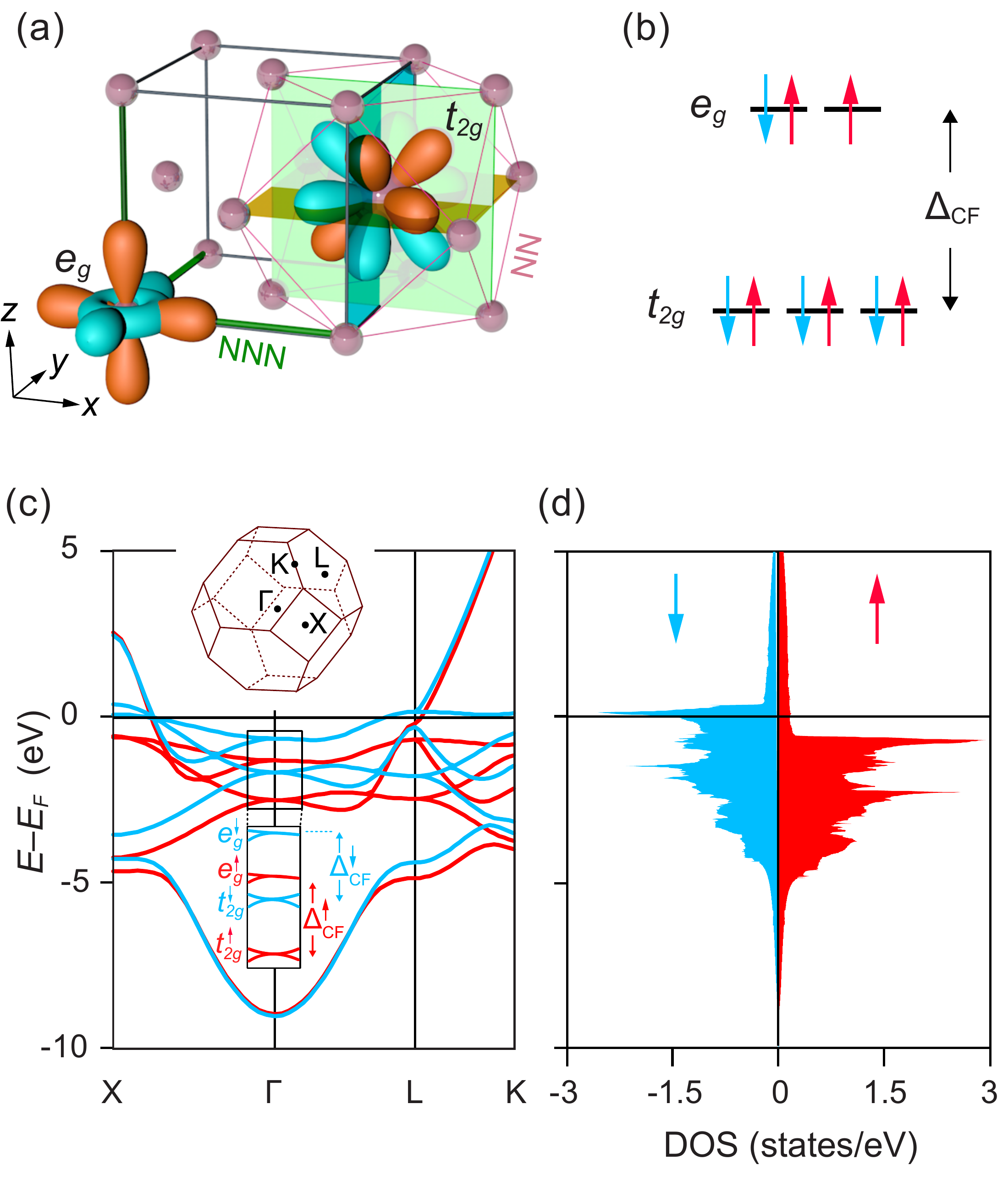}
	\caption{Schematic illustration of (a) the crystal structure of Ni and (b) its corresponding crystal field $\Delta_{\text{CF}}$, experienced by the $e_g$ and $t_{2g}$ orbitals. Each Ni is cuboctahedrally coordinated via $t_2g$ orbitals with its nearest neighbours (NNs), indicated with the purple lines in (a). NNN stands for the next nearest neighbour, also indicated by the thick green lines in (a). (c) and (d) The spin-resolved electronic band structure and density of states of Ni at ambient pressure, respectively. The inset in (c) is the corresponding Brillouin zone and high symmetry $k$-points. }
	\label{1}
\end{figure}

Here, we revisit this long-standing problem from a physicochemical perspective, taking into account the symmetry relations between the bonding orbitals and local magnetic moments. Systematically analysing the spin and orbital characters of the bonding states in Ni through first-principles calculations, we confirm the quenching of the FM ordering at high pressures and propose an inter-spin charge transfer mechanism as its driving force. The characteristic feature of this mechanism is a continuous transfer of charge from the spin majority channel to the spin minority channel, caused by an enhancement of crystal field splitting of Ni-$3d$ orbitals.

\begin{figure}[t]
	\centering
	\includegraphics[width=0.49\textwidth]{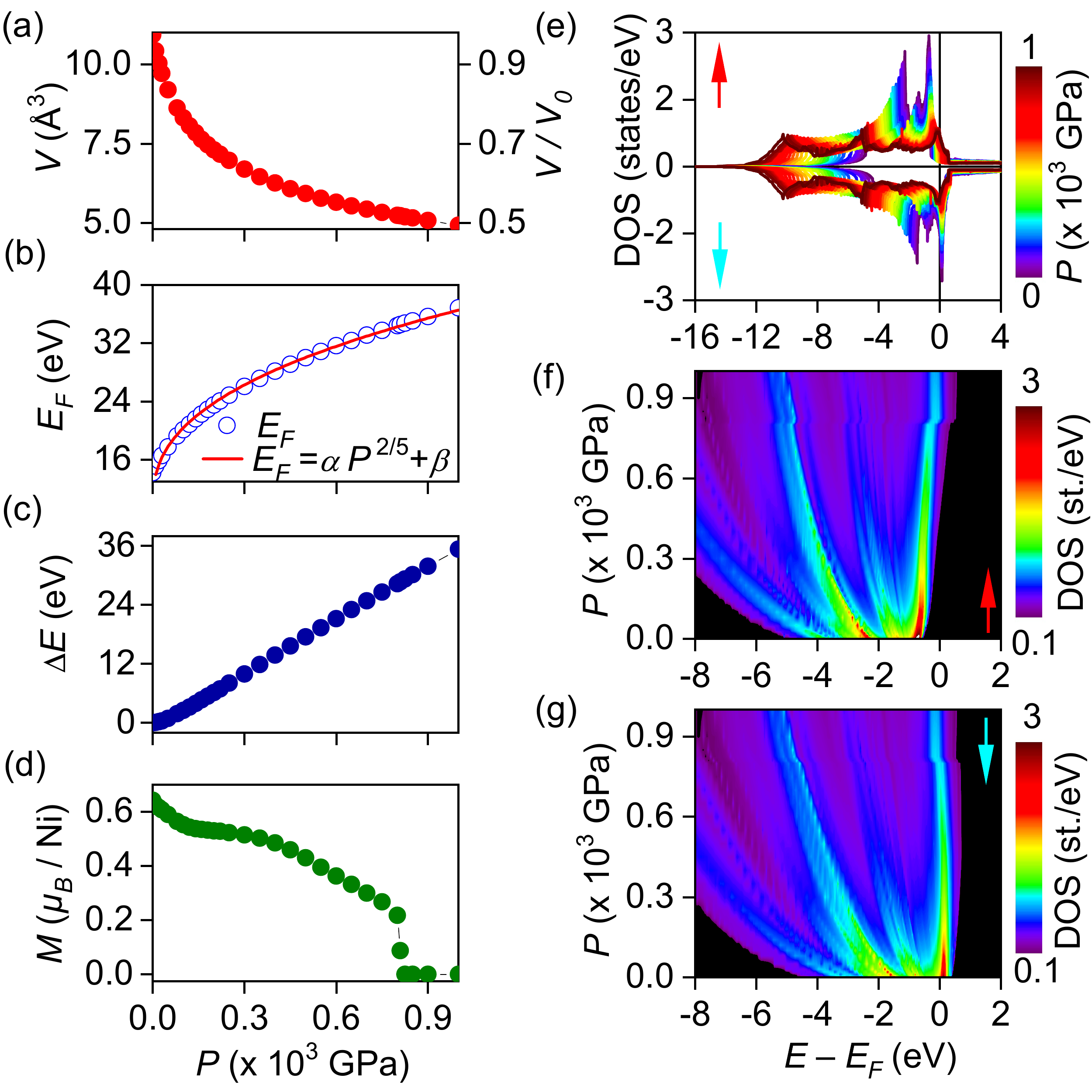}
	\caption{Pressure-dependence of (a) volume, (b) Fermi energy, (c) relative free energy $\Delta E=E-E_0$,  where $E_0$ is the free energy of Ni at the ambient pressure,  and (d) magnetization (M). The solid lines in (b) is the corresponding curve obtained from fitting the calculated results to the Sommerfeld's free electron model $E_F=\alpha P^{2/5}+\beta$, with 
%	$\alpha=1.693$ eV.(GPa)$^{-2/5}$ 
    $\alpha=6.740$ meV.Pa$^{-2/5}$ 
	and $\beta=9.698$ eV.  (e)-(g) Different representations of spin-resolved density of states as a function of pressure.}
	\label{2}
\end{figure} 

The electronic structure calculations are performed within DFT using the Perdew-Burke-Ernzerhof (PBE) exchange-correlation functional~\cite{Per1996} and ultrasoft pseudopotential as implemented in  \textsc{Quantum Espresso}~\cite{Gia2009}. The lattice parameter at ambient pressure is set to $a = 3.506$ \AA~\cite{Vce2003}. The effect of  $P$ is treated by optimising the volume of Ni lattice without breaking its FCC crystal symmetry using the Broyden-Fletcher-Goldfarb-Shanno algorithm~\cite{Fle2013}. The corresponding Brillouin zone (BZ) is sampled by a 60 $\times$ 60 $\times$ 60 $k$-mesh.

\begin{figure}[t]
	\centering
	\includegraphics[width=0.49\textwidth]{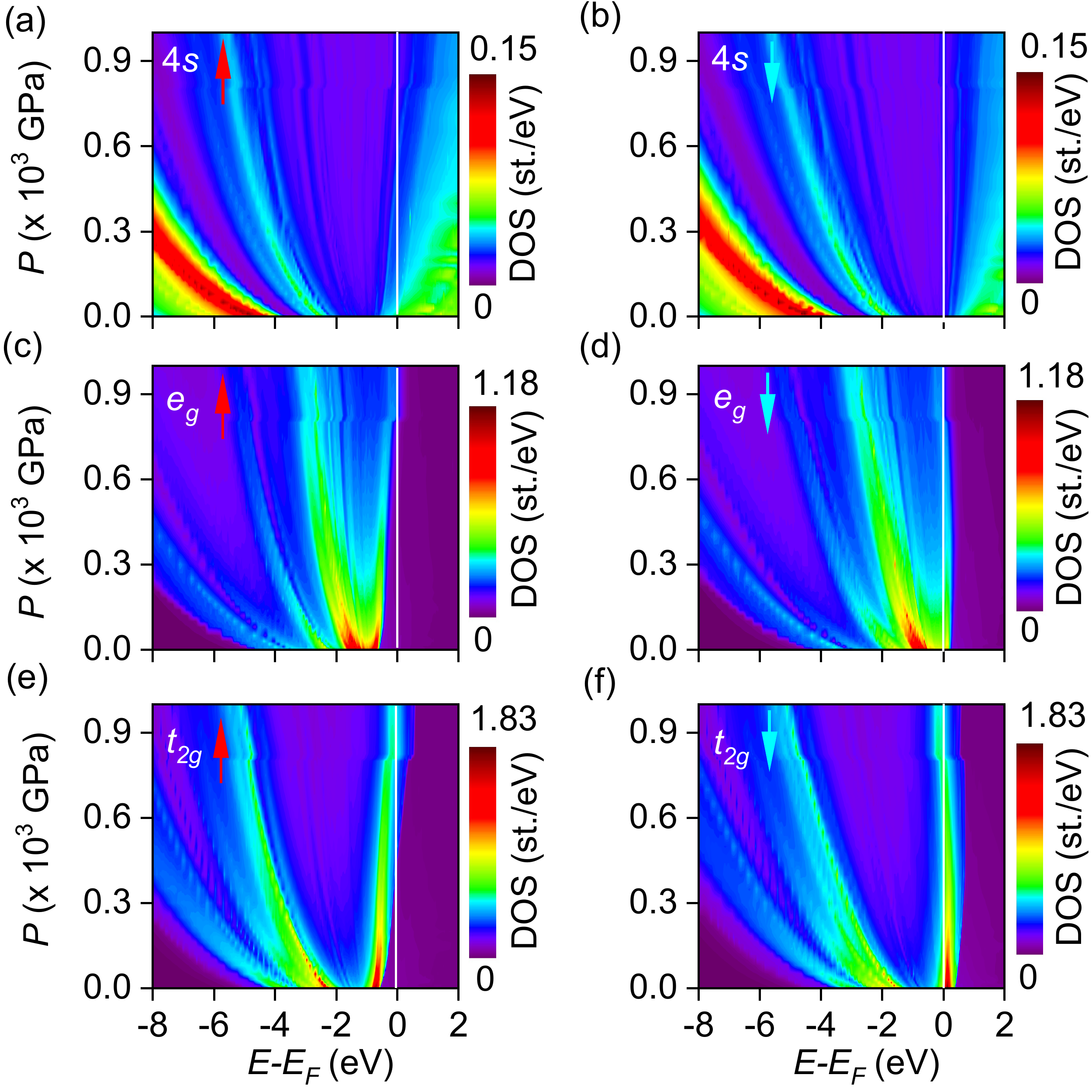}
	\caption{Heatmap plots of spin-projected density of states resulting from (a) $4s\uparrow$, (b) 4s $\downarrow$, (c) $e_g$ $\uparrow$, (d) $e_g$ $\downarrow$, (e) $t_{2g}$ $\uparrow$ and (f) $t_{2g}$ $\downarrow$ orbitals.}
	\label{3}
\end{figure} 

 The conventional unit cell of Ni is shown in Fig.~\ref{1} (a). This close-packed FCC structure allows each Ni to be coordinated with 12 neighbours~\cite{Bur1993,Dre2007}, thereby giving the whole structure the highest possible symmetry among all space groups (Fm-3m). Such cuboctahedral coordination gives rise to a cubic crystal field $\Delta_\text{CF}$, splitting the Ni-$3d$ orbitals into two subgroups $e_g$ composed of axially-oriented \{$d_{x^2-y^2}$ and $d_{z^2}$\} orbitals and $t_{2g}$ composed of diagonally-oriented \{$d_{xz}$, $d_{yz}$ and $d_{xy}$\} orbitals. 
 %Each subgroup is further split into a bonding and anti-bonding state with a momentum $k$-dependent gap, being Zero at the BZ centre $\Gamma$ point and reaching its maximum at the BZ boundary $X$ ($K$) for $t_{2g}$ ($e_g$). 
 As each nearest neighbour (NN) of Ni is diagonally coordinated, the $t_{2g}$ orbitals more effectively contribute to the Ni-Ni bonds (see Fig.~\ref{1} (b)). Accordingly, they tend to lie below the $e_g$ states and show a stronger energy dispersion. As a result, the $e_g$ states form a flat band at and near the Fermi level $E_F$, promoting strong \textit{e-e} Coulomb interaction. The system lowers this unfavourable energy term by undergoing time-reversal symmetry, causing a general spin splitting in its band structure. While this allows pushing the energy bands below $E_F$ in the spin majority channel ($\uparrow$), in the spin minority channel ($\downarrow$), there is still a significant contribution from $e_g$ flat bands, appearing as a sharp van hove peak in the density of states (DOS) at $E_F$, as shown in Fig.~\ref{1} (e) and (f). The resulting itinerant FM phase is found to hold a nominal $3d^{9}$ state~\cite{Blu2001} with a net saturated magnetisation of $M_s=0.6$ $\mu_B$/Ni, in good agreement with experiment~\cite{Ser2013}.

 To understand how $P$ affects the electronic structure, we have summarised the volumetric, magnetic and electronic properties of Ni as a function of pressure in Fig.~\ref{2}. As can be seen,  $V$, $E_F$  and free energy ($E$) respond monotonously to the $P$ variation (see Fig.~\ref{2} (a), (b) and (c), respectively), whereas  the magnetisation drops discontinuously  at the critical pressure $P_c=810$ GPa (see Fig~\ref{2} (d)). Particularly,  the increase in $E_F$ can be well described by Sommerfeld's power-law relation $E_F \propto P^{2/5}$, as shown in Fig.~\ref{2} (b). Such a disparity in magnetic and volumetric properties of Ni under $P$ is a strong indication of a spontaneous violation of the SC without structural symmetry breaking.  In other words, the $P$-induced inflation of the BZ is so strong that it can alone modify the dispersion of the energy bands, including the flat $e_g$ bands, such that the net DOS($E_F$) is no longer sufficient  to satisfy the SC for 
 FM ordering. %The question is how this happens microscopically.

The DOS plots shown in Fig.~\ref{2} (e) does support the scenario of the magnetisation collapse at high $P$s. This is more evident in the heatmap plots shown in Figs.~\ref{2} (f) and (g). As can be seen, the DOS near and at $E_F$ is abruptly perturbed at $P_c$. The trend of changes is, however, opposite between the two spin channels. While for spin-$\uparrow$ states, this appears as a sudden shift of DOS above $E_F$,  in the spin-$\downarrow$ channel, DOS shifts downward, suggesting a spontaneous \textit{inter-spin} charge transfer at $P_c$. Logically, this means a redistribution of charge, partly moving the $t^\uparrow_{2g}$ electrons used to form the NN Ni-Ni bonds to $e^\downarrow_g$ orbitals to enhance the next nearest neighbour (NNN) Ni-Ni bonds as they become increasingly shorter under $P$ (see Fig.~\ref{1} (c)). 

To examine this, we have calculated the spin- and orbital-projected DOS for a wide range of $P$s, see Fig.~\ref{3}. These calculations confirm that the pressure enhances the electronic dispersions of the bands throughout the whole energy spectrum and irrespective of their orbital characters. Interestingly, for $4s$ orbitals, this enhancement equally affects bonding and antibonding branches, causing the former (latter) to extend to lower (higher) energies, see Fig.~\ref{3} (a) and (b). As this implies a partial depletion of $4s$ charges near $E_F$, one may expect that part of the charge transfer mentioned above comes from the $4s$ electrons. However, the charge provided this way is not enough to fill the unoccupied $3d-\downarrow$ states (compare the DOS scales in Fig.~\ref{3}). Accordingly, the prime contributor for the expected band filling and its resulting magnetisation collapse is a charge transfer from $t^\uparrow_{2g}$ and $e^\uparrow_g$ to their corresponding spin-$\downarrow$ counterparts. This can indeed be seen in Fig.~\ref{3} (c)-(d). The most contrasting feature in these plots is a unilateral expansion of DOS towards lower energies under $P$. At and near $E_F$, the changes are minute but critical. Here one can notice a compensating behaviour between the two spin channels, appearing as an upward (downward) shift of the edge of DOS$^\uparrow$ (DOS$^\downarrow$). Together with the overall enhancement of band dispersions, this leads to a substantial decrease of DOS$^\downarrow(E_F)$. In other words, by applying $P$, the previously unoccupied $t^\downarrow_{2g}$ and $e^\downarrow_g$ states increasingly gain charge from their counterparts in the other spin channel. Remarkably at $P_c$, a sudden charge transfer takes place, abruptly pushing all the $t^\downarrow_{2g}$ and $e^\downarrow_g$ states below $E_F$ and quenching magnetisation entirely. 

\begin{figure}[t]
	\centering
	\includegraphics[width=0.48\textwidth]{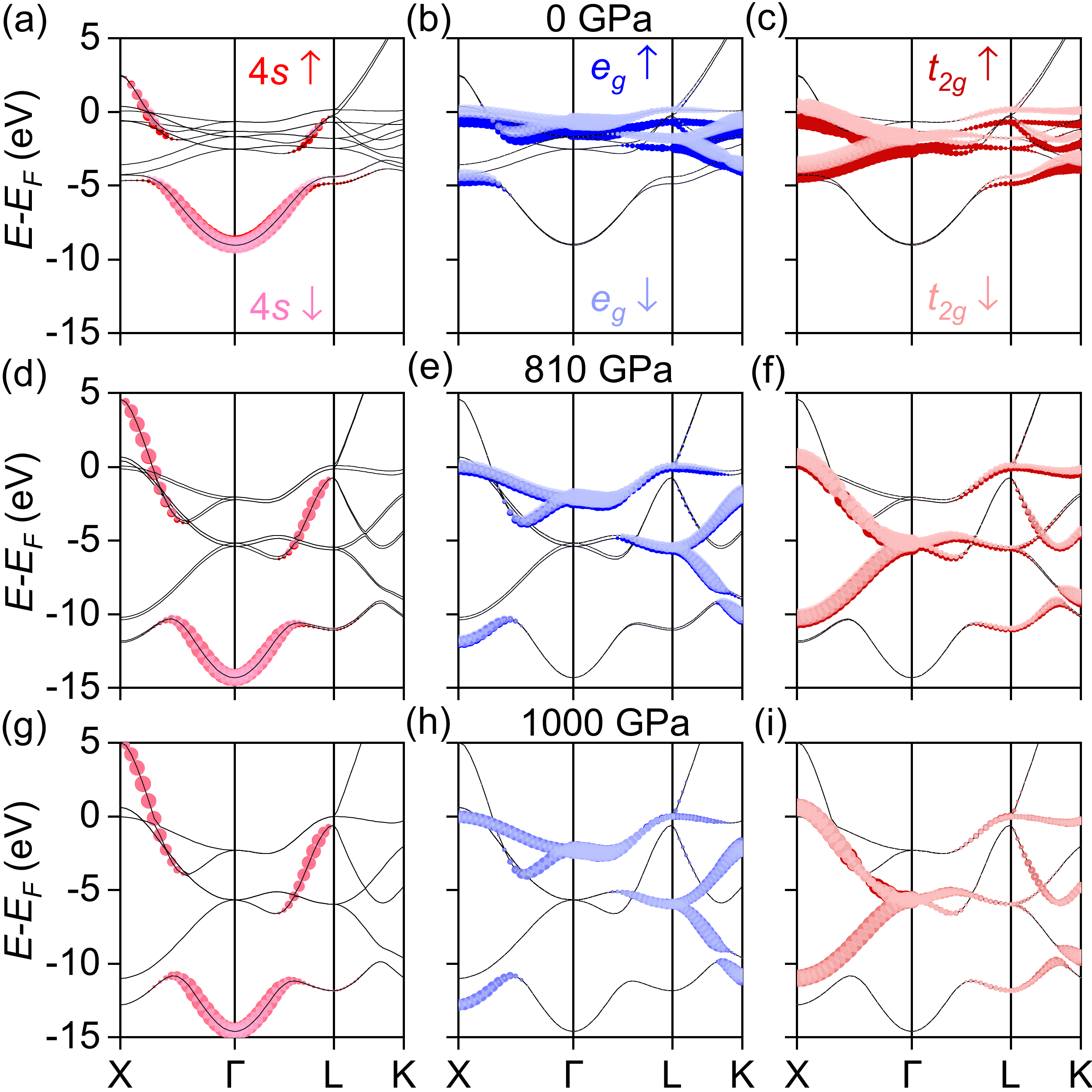}
	\caption{Spin- and  orbital-projected band structures of Ni calculated at the ambeit prssur (a)-(c), the critical pressure 810 GPa (d)-(f) and 1000 GPa (g)-(h).}
	\label{4}
\end{figure} 

To clarify this further, we have also calculated the spin- and orbital-projected band structures at three $P$s (0, 810 and 1000 GPa), representing low, critical and high $P$s, respectively see Fig.~\ref{4}. As is evident, the pressure enhances the dispersion of the bands, thereby weakening the overall Coulomb interaction to the extent that the bands become nearly spin-degenerate at high $P$s. The flatness of $e_g$ bands and localisation near $E_F$ at low $P$s can also be seen in this figure. Under $P$, they become increasingly dispersed, confirming strong electron hopping between the NNN Ni sites, as discussed earlier. One can also see a drastic enhancement of the bonding-antibonding gap of $t_{2g}$ states at $X$ point, followed by a sizable downward shift of the whole $t_{2g}$ bands, maintaining their band edge at $E_F$.   Such profound changes in the electronic structure clearly show how pressure acts against the SC through the cubic crystal field of Ni, enabling a complete collapse of magnetisation.  

 \begin{figure}[t]
 	\centering
 	\includegraphics[width=0.48\textwidth]{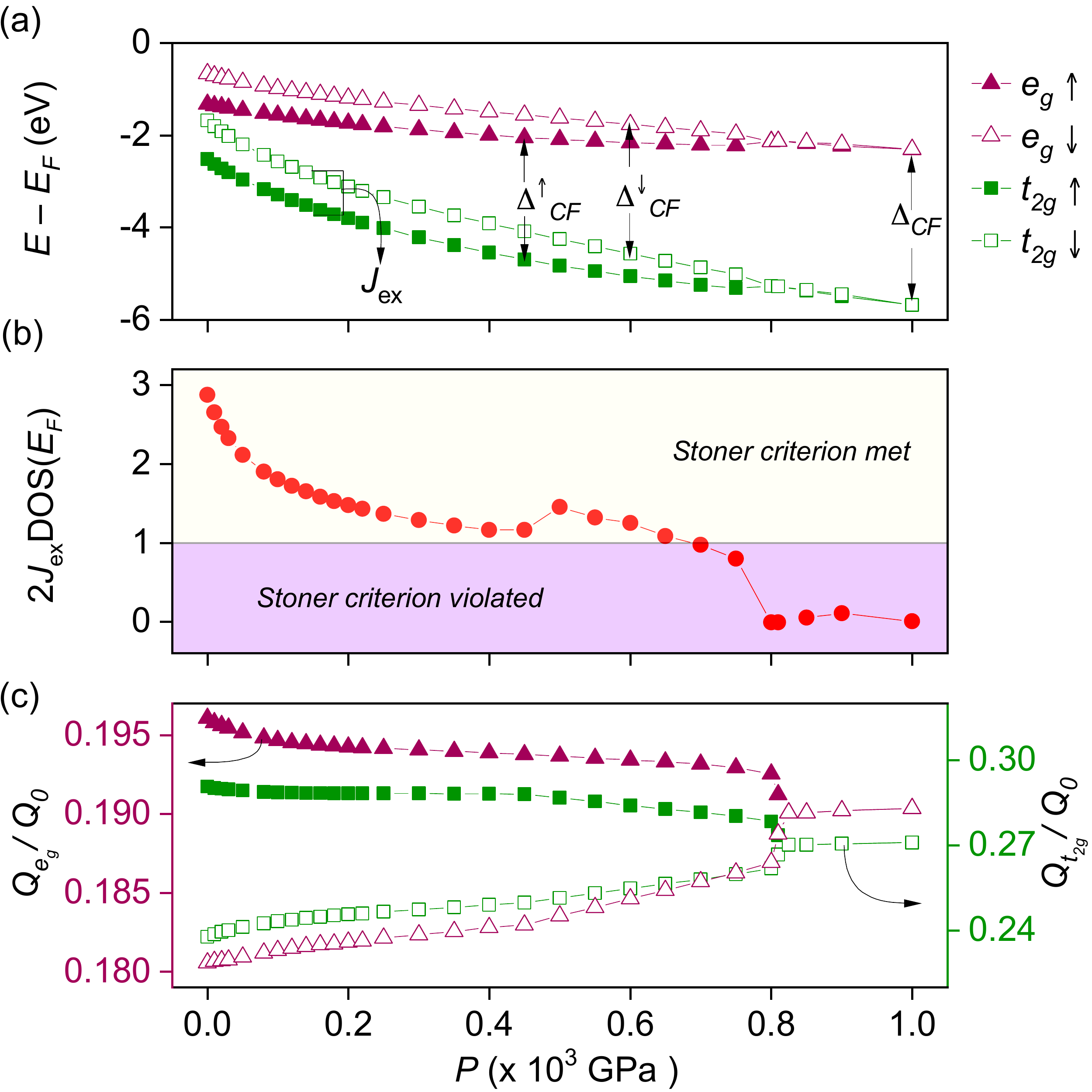}
 	\caption{(a) $P$-dependence of the crystal filed splitting $\Delta^{\uparrow\downarrow}_\text{CF}$ and the magnetic exchange splitting $J_\text{ex}$ between $e_g$ and $t_{2g}$ states. (b)The evolution of Stoner criterion as a function of $P$. (c) Normalized partial charges on Ni atom decomposed to  $e^\uparrow\downarrow_g$ and $t^\uparrow\downarrow_{2g}$ orbitals. $Q_0$ corresponds to the total charge at the Fermi level at the ambient pressure. }
 	\label{5}
 \end{figure} 

To demonstrate how $P$ violates the SC, we show in Fig.~\ref{5} (a) the $P$-dependence of $\Delta^\sigma_\text{CF}$ for both spin channels $\sigma=\uparrow$ and $\downarrow$. As can be seen, at low $P$s, $\Delta^\sigma_\text{CF}$ is small and comparable to the magnetic exchange coupling energy $J_\text{ex}$. This can be justified using a mean-field Ising model $\Delta E_\text{ex}=-J_\text{ex} \sum_{i=\text{NN}}\textbf{S}_0\cdot\textbf{S}_i=-12J_\text{ex}$, considering an FM ordering  between the twelve NNs of Ni. Within this regime, the obvious dominance of $\Delta E_\text{ex}$ does not allow the electrons to be delocalised, therefore, substantially suppressing $\Delta^\sigma_\text{CF}$. Increasing $P$ promotes the charge transfer, as explained above, and hence, monotonously enhances $\Delta^\sigma_\text{CF}$ against $\Delta E_\text{ex}$. Eventually it completely eliminates $J_\text{ex}$ and so becomes the dominant source of band splitting. Considering that $U=2J_\text{ex}$, we can readily show that the criterion $2J_\text{ex}\text{DOS}(E_F)> 1$ is violated once $P>P_c$, see Fig.~\ref{5}(b). This provides further evidence that the magnetisation collapse in Ni is due to the enhancement of $\Delta_\text{CF}$.

To quantify the rate and amount of charge transferred between the spin channels, we have performed L\"owdin charge population analysis using partial charge densities centred at Ni sites at various $P$s, see Fig.~\ref{5} (c). Our calculation reveals that for $P$s up to 400 GPa, the charge transfer is predominantly from $e_g^\uparrow$ to $e_g^\downarrow$ and $t_{2g}^\downarrow$. Within this range, the $t_{2g}^\uparrow$ charge, except for a minor drop at low $P$s, is nearly constant. This is consistent with our previous finding that the van hove singularities near $E_F$ are dominated by $e_g$ states. Beyond 400 GPa, the $t_{2g}^\uparrow$ population drops sharply and rapidly exceeds the $e_g^\uparrow$ charge drop. Consequently, the charge gained by  $e_g^\downarrow$ and $t_{2g}^\downarrow$  states boosts dramatically and grows linearly. 
%It is worth noting that such a linear behaviour can also be seen in the pressure-dependence of magnetisation above 400 GPa (Fig.~\ref{2} (d)). 
This implies that such massive compressive forces have shrunk the Ni lattice so much that its NNN Ni bonds are now as effectively involved in charge transfer as NN Ni bonds. Eventually, at $P_c$, the charge transferred from (to) $e_g^\uparrow$ and $t_{2g}^\uparrow$ ($e_g^\downarrow$ and $t_{2g}^\downarrow$) drops (jumps) abruptly and then remains constant. The system is now in a new equilibrium state with equal charges in both spin channels. As such, it becomes non-magnetic and remains so for $P\ge Pc$.

%It is worth mentioning a previous report has suggested that the competition of electron correlation and spin-orbit coupling (SOC) may result in magnetic anisotropy in Ni~\cite{Yan2001}, even though it naturally behaves as an itinerant ferromagnet~\cite{Don1994,Mor2013}. Some also have claimed that SOC alone plays a key role in the observed magnetisation drop in Ni under pressure. To examine the contribution of both electron correlation effects and SOC, we have performed additional calculations incorporating the onsite Hubbard U and SOC corrections. As can be seen in Supplementary Fig. S1, SOC has nearly no impact on electronic dispersion and magnetisation. We thus rule it out. Inclusion of the Hubbard U correction has no effect other than overestimating the magnetic moment at ambient pressure. Their combined effect also does not show any noticeable improvement. (see Supplementary Fig. S3). As such, we believe their contributions can be ignored.

In summary, we systematically investigated the origin of magnetisation quenching in the nickel crystal using first-principles calculations. The pressure was found to promote an inter-spin charge transfer, eventually equalising the charge population between $3d^\uparrow$ and $3d^\downarrow$ states. The driving force for this charge transfer was discussed to be the enhancement of the crystal field splitting between the $e_g$ and $t_{2g}$ bands, enabling the former to form chemical bonds between the next nearest neighbour Ni ions in the expense of partial depopulations of the latter. These findings shed new light on the behaviour of complex orbital manifolds subject to strong crystal fields and Coulombic interactions.

%\acknowledgments
MSB gratefully acknowledges the Research Infrastructures at the University of Manchester for allocations on the CSF3 high performance computing facilities. AA is grateful to Dr. D. K. Shukla for his encouraging supports. 

%\bigskip
%\bibliography{Nickel}

\end{document}